# Reconfigurable classifier based on spin torque driven magnetization switching in electrically connected magnetic tunnel junctions

A. López[1,2], D. Costa[3], T. Böhnert[3], P. P. Freitas[3], R. Ferreira[3], I. Barbero[1], J. Camarero[2,5], C. León[1,4], J. Grollier[6], M. Romera[1,4*]

[1]GFMC, Departamento de Física de Materiales, Universidad Complutense, Madrid, Spain.
[2]IMDEA Nanociencia, C/Faraday 9, 28049 Madrid, Spain.
[3]International Iberian Nanotechnology Laboratory (INL), Braga, Portugal.
[4]Unidad Asociada UCM/CSIC, Lab. de Heteroestructuras con Aplicación en Espintrónica, 28049 Madrid, Spain.
[5]Departamento de Física de la Materia Condensada and Departamento de Física Aplicada, IFIMAC and Instituto Nicolás Cabrera, Universidad Autónoma de Madrid, 28049 Madrid, Spain.
[6]Unité Mixte de Physique CNRS, Thales, Université Paris-Sud, Université Paris-Saclay, Palaiseau, France.

*miromera@ucm.es

A promising branch of neuromorphic computing aims to perform cognitive operations in hardware leveraging the physics of efficient and well-established nano-devices. In this work, we present a reconfigurable classifier based on a network of electrically connected magnetic tunnel junctions that categorizes information encoded in the amplitude of input currents through the spin torque driven magnetization switching output configuration. The network can be trained to classify new data by adjusting additional programming currents applied selectively to the junctions. We experimentally demonstrate that a network composed of three magnetic tunnel junctions can learn to classify spoken vowels with a recognition rate that surpasses the performance of software multilayer neural networks with the same number of trained parameters in this task. These results, obtained with the same nano-devices and working principle employed in industrial spin-transfer torque magnetic random-access memories (STT-MRAM), constitute an important step towards the development of large-scale neuromorphic networks based on well-established technology.

## I. INTRODUCTION

The advancement of neuromorphic hardware holds the potential to address cognitive computing tasks, like pattern recognition, with significantly reduced energy consumption [1-3]. This requires dense networks of low power processing units able to learn to classify information into categories. In recent years, spintronic nano-devices have shown great potential as building blocks in hardware neural networks [4-21] thanks to their multifunctionality, nano-scale dimensions, room temperature operation, high endurance and compatibility with complementary metal–oxide–semiconductor (CMOS) technology. Many of these proposals are based on magnetic tunnel junctions (MTJs) in which a spin polarized current manipulates the magnetization of the free layer via the spin transfer torque effect [4-15]. These junctions can exhibit different functionalities of interest for neuromorphic computing depending on the regime of magnetization dynamics activated [4,12]. For instance, in spin torque oscillators the magnetization of the free layer is driven into a regime of self-sustained precession, which gives rise to the emission of a microwave signal [22]. It was recently shown that a single spin torque oscillator can perform spoken-digit recognition thanks to the combination of its high non-linearity, stability and aspect to noise ratio [5]. Moreover, a network of four coupled spin torque oscillators has learnt to classify spoken vowels encoded in the frequencies of microwave signals [6]. In this approach, the synchronization configuration output by the network in response to the microwave signals is used to categorize the input data. In another study, an assembly of three spintronic oscillators has shown ability to bind events through their mutual synchronization to classify temporal sequences [9]. More recently, a multilayer neural network based on spin torque oscillators that communicate through radio frequency signals have been reported [15].

Superparamagnetic tunnel junctions are another spintronic device with great potential for neuromorphic and probabilistic computing [7,23-25], including true random number generators [23] and probabilistic spin logic [24]. These junctions are characterized by a superparamagnetic free layer whose magnetization fluctuates stochastically between parallel (low resistance) and antiparallel (high resistance) states with equal probability due to thermal noise. In the context of neuromorphic computing, their stochastic switching behavior have been used to map the probabilistic spiking nature of biological neurons [26], which is believed to be important for their reduced energy consumption. The switching probability of these junctions can be tuned with an applied current, leading to spiking generation with tunable spiking rate which resembles the behavior of biological neurons [13,27]. Assemblies of superparamagnetic tunnel junctions have leveraged these features to implement neuron functionalities such as rate coding and population coding [7].

However, the most mature and well-established spintronic technology today is STT-MRAM memory, which relies on magnetic tunnel junctions in which the spin transfer torque effect leads to the free layer magnetization switching between two stable states (parallel and antiparallel) characterized by a different resistance [28]. Industrial STT-MRAM memories consist of arrays comprising hundreds of millions of these spin torque driven magnetization switching nano-devices (from now on, spin torque switching devices), fabricated and electrically interconnected on top of CMOS chips. In those arrays, the magnetic configuration of the junctions can be written and read efficiently and fast with dc currents or current pulses through the spin transfer torque and the magnetoresistance effect respectively. The realization of complex computing tasks with spin torque switching nano-devices as those used in STT-MRAM memories holds the potential of a simple and fast integration on a large scale, making it a very attractive technological prospect. However, so far the use of spin torque switching devices in neuromorphic networks have been investigated mainly as synaptic elements, which are usually combined with other kind of (software) non-linear processing units to perform classification tasks. For instance, the stochasticity of the switching process between the parallel and antiparallel states has led these junctions to be theoretically proposed as stochastic binary synapses [8,29], and chains of electrically connected switching devices have been investigated as spintronic memristors [30] to create quantized synaptic weights [31]. More recently, crossbar arrays of these junctions have been used to implement synaptic weights and to perform in an analogue manner multiply–accumulate (MAC) operations in a two-layer neural network for in-memory computing [32]. Nevertheless, so far spin torque switching nano-devices have not been employed experimentally as interconnected processing units in neuromorphic networks capable of learning to perform a classification task.

In this work, we apply a learning procedure initially proposed by a network of spintronic oscillators for pattern recognition [6] to a more mature and simpler technology. We show that a chain of electrically connected spin torque switching nano-devices identical to those used in STT-MRAM memories can learn to categorize data through its switching configuration (see schematics in Figure 1a). In this computing scheme, inputs are dc current signals (input currents, $I_{input} = I_A$ and $I_B$) encoding information in their amplitudes and the output is the network switching configuration. Learning is implemented through additional dc current signals which can be applied selectively through the devices simultaneously to the input signals. These signals, which we refer to as control currents ($I_{C1}, I_{C2}, I_{C3}$, see Figure 1a), are used to modify the threshold input current above which spin torque driven magnetization switching is induced in each device. We refer to this threshold input current as the effective critical input current, $I_{input}^{th}$. We show that a network based on three spin torque switching nano-devices can learn to classify seven spoken vowels with a recognition rate of 96%, which is above the rate that can be obtained with software neural networks with the same number of trained parameters in this task.

## II. EXPERIMENTAL RESULTS

We electrically connect three spin torque switching nano-devices as described in the schematic of the experimental set up shown in Figure 1a and in Figure S1 in Supplemental Material [34]. The devices are in-plane magnetized magnetic tunnel junctions with a structure of 100 $Al_2O_3$/3 Ta/30 CuN/5 Ta/17 $Pt_{38}Mn_{62}$/2 $CoFe_{30}$/0.85 Ru/2.6 $CoFe_{40}B_{20}$/0.85 MgO/1.4 $CoFe_{40}B_{20}$/10 Ru/150 Cu/30 Ru (thicknesses in nm) [33]. Upon deposition, the wafers were annealed for 2 h at 330 °C and cooled down under a magnetic field of 1 T along the easy axis defined during deposition. Circular devices with a diameter of 150 nm were then patterned combining standard lithography and etching techniques. The devices resistance in the parallel state is close to 1100 Ω and the magneto-resistance is about 120% at room temperature (see Figure S2 in Supplemental Material [34]). The three magnetic tunnel junctions are electrically connected using wire bonding. In our convention, positive current corresponds to electrons flowing from the polarizer to the free layer favoring the parallel state. An in-plane magnetic field of $\mu_0 H = 6$ mT is applied along the easy axis during electrical measurements to facilitate spin torque driven magnetization switching between the parallel and antiparallel states. Current signals of two different categories are used in the reconfigurable classifier. On one hand, current signals encoding information in their amplitudes are presented as inputs to the network ($I_{input}$) and flow across all the junctions. These are the inputs to be classified by the network. On the other hand, control currents can be applied selectively through each device ($I_{C1}$, $I_{C2}$, $I_{C3}$) to program (train) the network response to the inputs. Thus, the actual current flowing through each device when an input is applied is $I_{MTJ1}=I_{C1} + I_{input}$, $I_{MTJ2}= I_{C2}+ I_{input}$ and $I_{MTJ3}= I_{C3}+ I_{input}$, where $I_{MTJi}$ corresponds to the current flowing through the $i$th junction (Figure 1a). Control currents can be modified during training and are kept fixed at inference when inputs are applied. In our experimental set up we use three different current sources to apply three dc current signals to the circuit ($I_{dc1}$, $I_{dc2}$, $I_{dc3}$), as shown in Figure 1a. These three applied dc current signals include the information of both input and control currents. For instance, the first current source applies a dc current signal $I_{dc1}$ that includes the input signal and the control current of the first junction ($I_{dc1}= I_{C1} + I_{input}$). Given the network electrical configuration (see Fig. 1a) this signal flows through all the junctions. While the input signal is meant to flow across all the junctions, the control current $I_{C1}$ should be applied only to the first junction (MTJ1). This is achieved by using current sources 2 and 3 to regulate the control currents of MTJ2 and MTJ3 respectively. More specifically, the currents applied by sources 2 and 3 are defined by the relations $I_{dc2}= I_{C2} -I_{C1}$ and $I_{dc3}= I_{C3} -I_{C2}$. We note that this configuration allows an independent control of the current that flows through each device, given by $I_{MTJ1}=I_{dc1}$, $I_{MTJ2}=I_{dc1}+I_{dc2}$ and $I_{MTJ3}=I_{dc1}+I_{dc2}+I_{dc3}$.

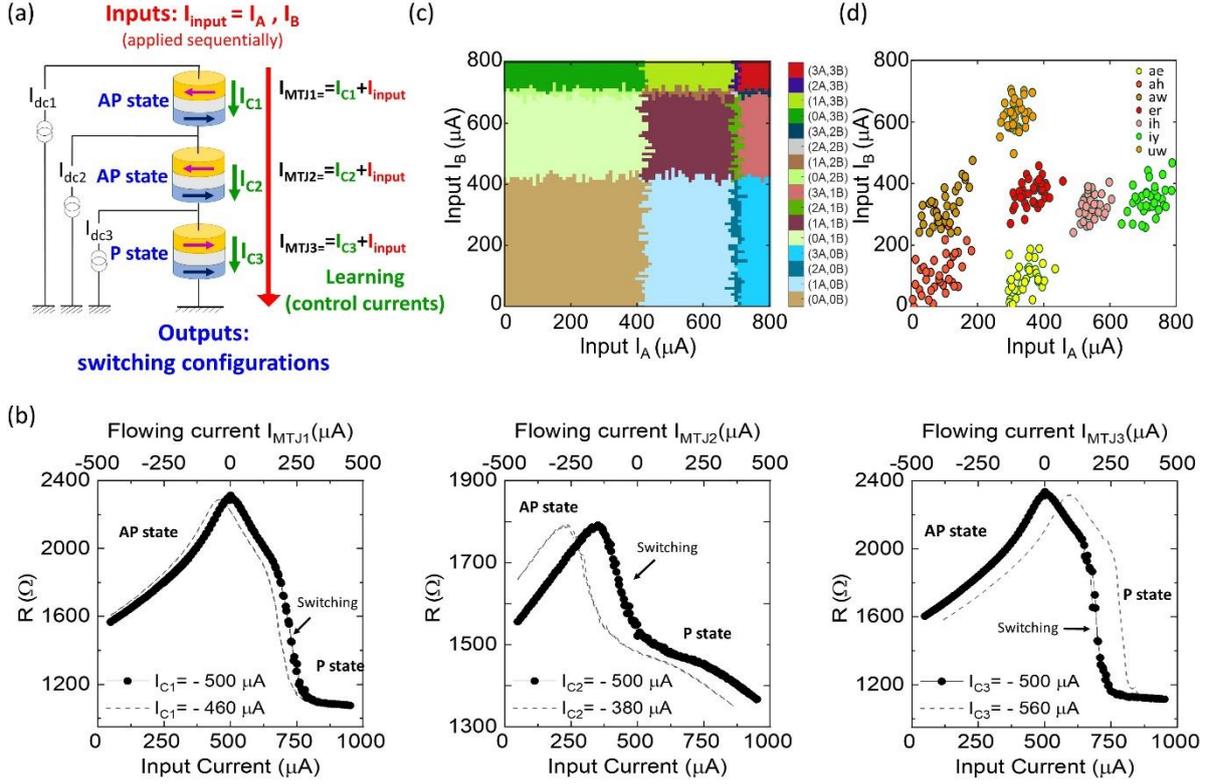

Figure 1: (a) Schematic of the experimental set up with three electrically connected spin torque switching devices. Two input dc current signals encoding information in their amplitudes ($I_{input}$) are applied sequentially to the network and flow across the three junctions. Control currents $I_{C1}$, $I_{C2}$ and $I_{C3}$ can be applied selectively to each device simultaneously to the input signals. The resistance of each junction is recorded with a voltmeter (not shown). (b) Resistance vs input current (bottom axis) for the three magnetic tunnel junctions with control currents $I_{C1} = I_{C2} = I_{C3} = -500$ µA (black symbols) and $I_{C1} = -460$ µA, $I_{C2} = -380$ µA, $I_{C3} = -560$ µA (dashed lines). The threshold input current at which the magnetization of each device switches can be controlled independently with the control currents. The top axis indicates the current flowing through each device for the curve with $I_{C1} = I_{C2} = I_{C3} = -500$ µA (black symbols). (c) Experimental switching map as a function of the amplitude of inputs $I_A$ and $I_B$, obtained with the same control current $I^j_{Ci} = -500$ µA for the three devices $i = 1, 2, 3$ and the two inputs $j = I_A, I_B$. Each color corresponds to a different output switching configuration. The map is obtained out of 160 experimental RvsI curves. (d) Inputs $I_A$, $I_B$ applied to the network encoding information of seven spoken vowels. Each color corresponds to a different vowel and each data point to a different speaker.

The control currents of the three devices are first set to the same value $I_{C1} = I_{C2} = I_{C3} = -500$ µA. Under these conditions of magnetic field and applied current the junctions are in the antiparallel state. We then apply an increasing input current $I_{input}$ across the chain of junctions in addition to the fixed control currents. Black symbols in Figure 1b show the resistance of the three junctions as a function of the input current $I_{input}$ (bottom axis). For clarity, the current flowing through each device is shown in the top axis. Figure 1b (black symbols) shows that the increasing input current induces spin torque driven magnetization switching from the antiparallel (AP state) to the parallel state (P state) in the three devices at different values of the input current ($I^{th1}_{input} = 724$ µA, $I^{th2}_{input} = 434$ µA and $I^{th3}_{input} = 696$ µA for junctions 1, 2 and 3 respectively) due to typical device-to-device variability. In consequence, the network naturally categorizes the input amplitude range through the number of magnetization switching events it induces. For instance, when three switching events occur, it indicates an input current exceeding 724 µA. Similarly, the absence of switching events suggests an input current below 434 µA. In the following we show that the number of switching events induced in the network can be used to recognize data, despite the typical device-to-device variability of current nano-fabrication techniques and the intrinsic stochasticity of spin torque driven magnetization switching. Importantly, by adjusting the control currents we can independently modify the effective critical input current of switching of each device. This is illustrated in Figure 1b (dashed lines), where the resistance of the three junctions is plotted as a function of the input current $I_{input}$ when the control currents are adjusted to $I_{C1}$ = - 460 µA, $I_{C2}$ = - 380 µA, $I_{C3}$ = - 560 µA. It shows that we can shift the threshold input current of switching in each junction independently by finely tuning the control currents. This will be used in the following to implement learning.

We now leverage this behavior to classify information encoded in the amplitude of two input dc current signals ($I_{input} = I_A$, $I_B$) applied sequentially to the network. First, input $I_A$ is applied, which can lead, depending on the conditions, to spin torque driven magnetization switching in one or more junctions. The resistance of the three junctions is monitored and analyzed in real time. The number of switching events detected is recorded in the computer. Then the input current is set to zero (so the junction is reset to the antiparallel state) and the procedure is repeated upon applying the second input $I_B$. The output of the network is the *switching configuration,* for instance [2A, 3B] if two switching events are induced by input $I_A$ and three switching events are induced by input $I_B$. Figure 1c shows the 16 different switching configurations experimentally obtained with this network as a function of two input currents $I_A$ and $I_B$, using a control current of $I^{j}_{Ci}$ = -500 µA for the three devices ($i$=1, 2, 3) and the two inputs ( $j = I_A$ and $I_B$).

We now test the network performance classifying real data into categories. For this we use a dataset of seven spoken vowels pronounced by 37 female speakers [6] (see S3 in Supplemental Material [34]), in which each vowel is characterized by 12 characteristic frequencies called formants. We encode the spoken vowels in the amplitude of two dc inputs $I_A$ and $I_B$ obtained through linear combinations of the characteristic frequencies as described in Supplemental Material S3 [34]. The inputs $I_A$ and $I_B$ presented to the network are shown in Figure 1d, in which each color corresponds to a different vowel and each datapoint to a different speaker. To achieve high classification rates, all the datapoints corresponding to each vowel in Fig. 1d should induce the same switching configuration in the network and therefore fall in the same color region in Fig. 1c.

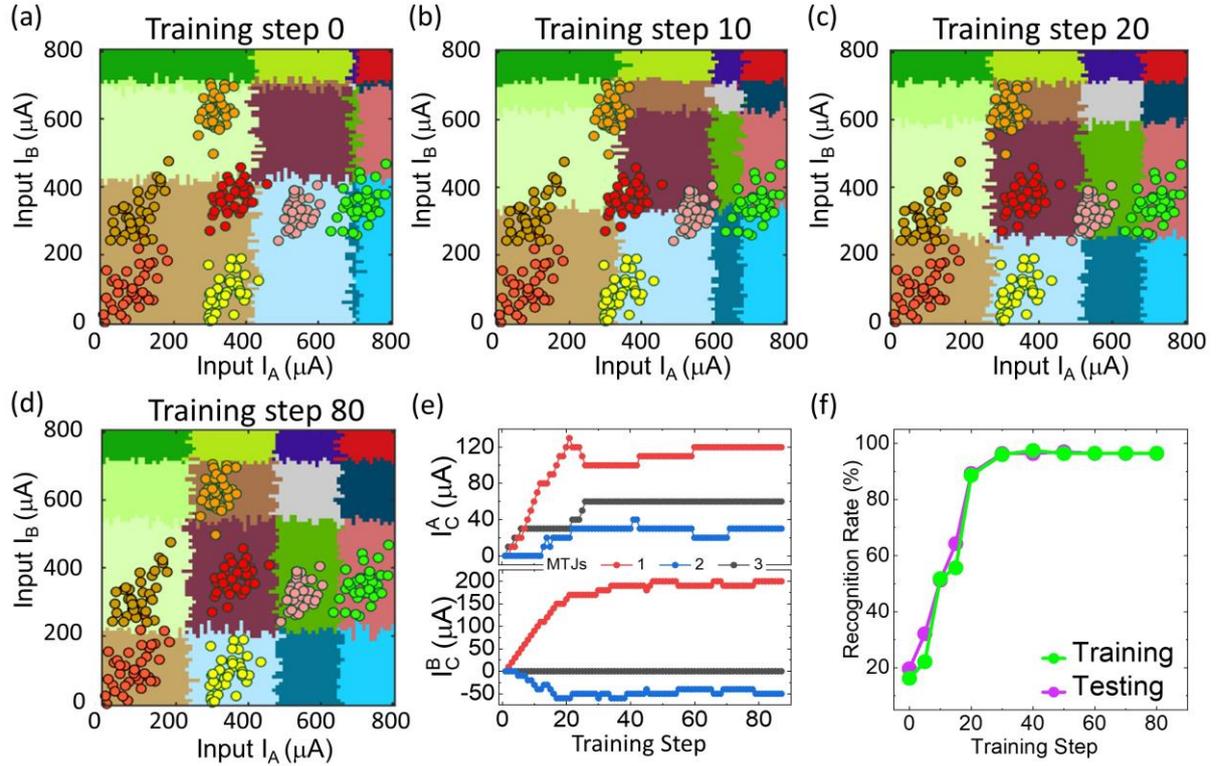

Figure 2: (a-d) Experimental switching maps as a function of the input dc currents $I_A$ and $I_B$, at different steps of the training process. (a) step 0; (b) step 10; (c) step 20 and (d) step 80. The color data points correspond to the inputs applied to the chain of junctions encoding information of spoken vowels pronounced by different speakers. Different colors represent different vowels. (e) Control currents applied to the three junctions upon applying input $I_A$ (top) and input $I_B$ (bottom), as a function of the number of training steps. Each color corresponds to a different junction. (f) Recognition rate obtained for the training and the testing data subsets as a function of the number of training steps.

The result of directly applying the inputs of Fig. 1d to the network is shown in Fig. 2a. As it can be seen, initially the network does not classify these inputs correctly. It needs to be trained to properly separate the different spoken vowels into categories. The switching regions in the experimental map of Fig. 2a are delimited by the effective critical input current at which the magnetization of each device switches ($I^j_{\text{th}i}$ for the three devices $i = 1, 2, 3$ and the two inputs $j = I_A, I_B$). Interestingly, we can independently modify the effective critical input current of switching of each device by tuning the control currents, as shown in Figure 1b (dashed lines). To train the network to classify the inputs of Figure 1d, the control currents should be adjusted so that each switching configuration in the experimental map approaches the input vowel that it is expected to classify. For this, we implement an automatic real-time supervised learning process [6]. In each learning step, a randomly chosen input of each category (i.e. seven inputs in total) is applied to the junctions. The switching configuration output by the network in response to each input is analyzed automatically in real time and stored in the computer until the seven vowels are applied. If the output switching configurations fails to correctly classify the inputs, the control currents $I^j_{Ci}$ of the three devices ($i= 1, 2, 3$) for the two inputs ($j = I_A, I_B$) are automatically modified either by +10 uA or -10 uA towards reducing misclassification errors (see S4 and S5 in Supplemental Material [34] for more details). The result of applying this procedure can be seen in Figure 2, which shows experimental switching maps at different steps of the training procedure (Figure 2a-2d), the evolution of the control currents $I^j_{Ci}$ (Figure 2e) and the recognition rate (Figure 2f) as a function of the number of training steps. Initially, the classification rate of the network is very low (Fig. 2a and 2f). In consequence, the learning algorithm modifies the control currents (Figure 2e) so that the output switching configurations in the experimental map are shifted approaching the clouds of datapoints they are meant to classify (Fig. 2b and 2c) and the recognition rate increases progressively (Fig. 2f). After around 30 steps the recognition rate reaches a high value and the algorithm ceases to make significant modifications to the control currents (Figure 2e and 2f). This indicates that the control currents have reached optimum values and the network has learnt to correctly classify the inputs. Indeed, at this stage the different input categories are well contained within their respective switching configurations (Fig. 2d). The training process is performed using only 80% of the database. When the training process ends, the resting 20% of the data is used to test the recognition of the network to new inputs. Figure 2f shows the recognition rate achieved by the network with the training and testing datasets as a function of the number of training steps. At the end of the training procedure the recognition rate is 96.6% for the subset of the database used for training, and 96.4% for the testing subset. We can compare this value with the performance of a software multilayer neural network with a similar number of trained parameters on the same task [6]. Figure 3 shows the recognition rate obtained by two multilayer perceptrons based on 27 and 47 trained parameters respectively trained through backpropagation in this classification task as a function of the number of training steps (see S6 in Supplemental Material [32] for details). The multilayer perceptron with 47 trained parameters reaches a recognition rate of 95%. Our network achieves a higher recognition rate despite being based only on 3 devices and 32 trained parameters (26 coefficients for the linear combinations to transform formants to inputs -see S3 in Supplemental Material [34]- and 6 control currents). These results show that the physics and working principle of spintronic nano-devices used in STT-MRAM memories can be leveraged to efficiently compute and categorize data.

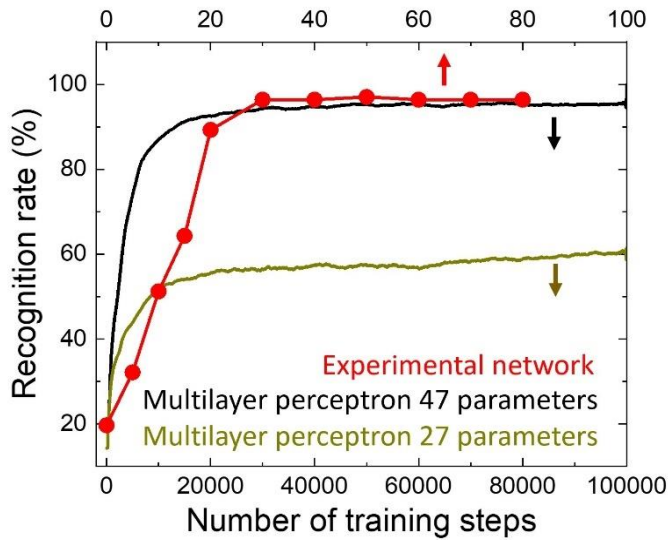

Figure 3: Evolution of the recognition rate obtained with the testing subset using the experimental network of spintronic switching nano-devices based on 32 trained parameters (red symbols) and a multilayer perceptron with 27 (green line) and 47 (black line) trained parameters as a function of the number of training steps. Note that in both the experimental and software networks the learning rate has been tuned to maximize the recognition rate and not to minimize the number of training steps required to train the network (see Supplemental Material S6 [34] for details). The experimental network and the multilayer perceptron were both trained with the same training subset.

### III. INTEGRATION

In this section we intend to provide insights into how the small proof-of-concept that we have developed in this study can be scaled up and integrated at an industrial level, and to discuss the effect of scaling up the network in training and in the recognition capacity.

The proposed reconfigurable classifier can classify $(N+1)^2$ categories encoded in two input signals into different switching configurations, where N is the number of devices in the network. In the future, the number of categories that can be classified and the complexity of the computing tasks that can be performed can be enhanced by using a larger number of input signals and scaling up the number of devices in the network. This computing scheme is indeed advantageous towards integrating and scaling up these networks [11], as hundreds of millions of these junctions with lateral dimensions of few nanometers [35] are fabricated and interconnected electrically on top of industrial CMOS chips for STT-MRAM memories. The maximum number of devices that can be contained in a chain of junctions maintaining a high recognition rate will be defined by (i) the maximum current that can flow across an MTJ without reaching the breakdown voltage, and (ii) the minimum difference that the effective critical input current of different devices can have without decreasing the recognition performance, which will in turn be limited by the intrinsic stochasticity of spin torque driven magnetization switching. In future networks, inputs and control dc current signals can be replaced by short current pulses encoding information in their amplitude and/or length. This will enhance energy efficiency and also increase the operational window of MTJs, whose barriers can sustain short pulses of very large current (and voltage) amplitude, well above the DC breakdown value, to switch the free layer magnetization. To scale up future networks further, multiple inputs will be processed simultaneously by different chains of junctions working in parallel, increasing in turn the computing speed.

It is worth noting that the classification regions obtained with this scheme can have different sizes and shapes (see for instance elongated green region [2A, 1B] and square dark brown region [1A, 1B] in Fig. 1c), allowing classification of asymmetric data. This can be designed before fabrication (since the critical current of magnetization switching can be defined through the devices diameter and aspect ratio) or after it, through control currents used to finely adjust the size and shape of each switching configuration in the experimental map, allowing learning.

The process through which an integrated large-scale reconfigurable classifier based on spin torque switching nano-devices classifies input data into categories is schematically shown in Figure 4 for the example of spoken vowels classification, including peripheral circuits.

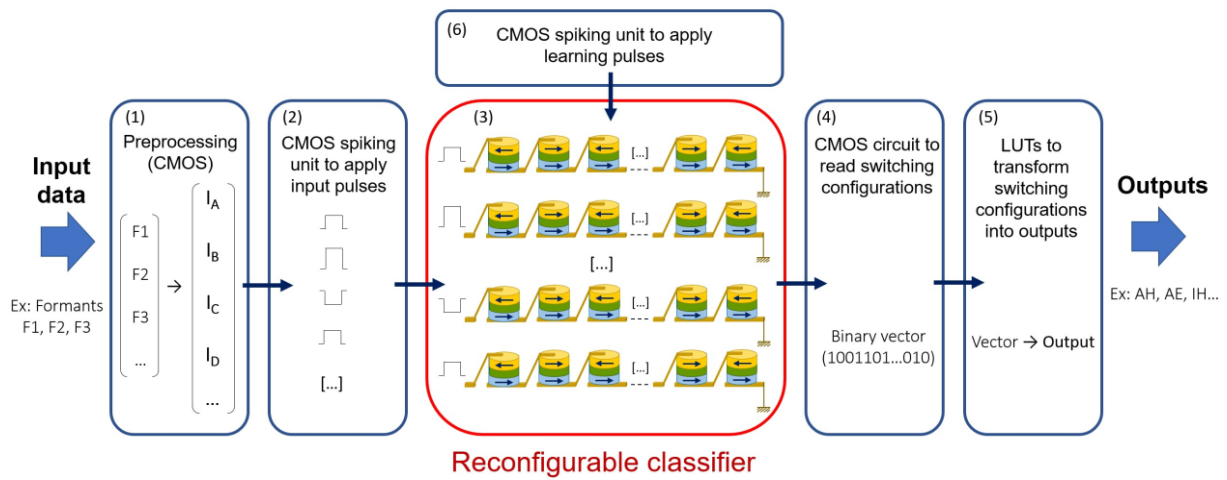

Figure 4. Schematic of the process through which the reconfigurable classifier can categorize input data. In the example of spoken vowel classification, inputs are the characteristic frequencies of spoken vowels (formants), and outputs are the associated vowels (see text).

Firstly, the original input data to classify (in this example, the formants) is transformed to current amplitudes ($I_A$, $I_B$, $I_C$, …) in the working range of the switching devices through simple multiplication operations or a low-frequency linear combination that can be implemented in CMOS [2, 36] or with programmable resistive devices [37] ((1) in Fig. 4). Then, current pulses encoding information in their amplitudes or lengths are applied simultaneously to the network of switching devices by CMOS spiking neurons [38] ((2) in Fig. 4) and processed in parallel by different chains of devices (as many chains as inputs, (3) in Fig. 4). Switching devices can be designed and engineered with different diameters and aspect ratios so they naturally have different critical currents. This approach minimizes energy consumption since the amplitude of learning pulses can remain low. Pulse amplitudes of the order of tens of µA are used in small junctions [39]. We note that endurance values of MTJs in STT-MRAM memories are in the order of $10^{14}$ to $10^{16}$ write cycles or more. The resistance variation associated to each spin torque driven magnetization switching is detected as in standard STT-MRAM memories. This is typically done by a simple circuit with 1 transistor per unit cell, although there are advanced circuits with a reduced number of transistors [40] ((4) in Fig. 4). The output of this detection circuit can be represented as a binary vector of dimension N (100111101001….1), where 1 represents switching, 0 represents no switching, and N is the number of switching devices. This binary vector is uniquely associated to a specific output category with lookup tables (LUTs) [41], which identify the category (output) to which the pattern (input) belongs ((5) in Fig.4). The final output of the system is the vowel that the classifier associates to the input. To implement learning, additional pulses can be applied simultaneously to the input pulses. This can be done by CMOS spiking units [38] directly connected to each device of the chain ((6) in Fig. 4) through a cross geometry. Learning and input current pulses are sent through opposite arms of the cross so each learning pulse only flows through one device. Learning pulses have different amplitudes and can have either the same or opposite polarity as input pulses. The amplitude of learning pulses is low, well below the critical current of magnetization switching, so a perfect synchronization with input pulses is not required. In case the use of devices with identical nominal diameter is preferred in industrial fabrication processes, the amplitude of learning pulses will be larger. In such case, the classifier can be designed so that learning pulses and input pulses have opposite polarity, so that the maximum input current that can be applied (and thus the maximum number of devices in each chain) is not affected.

### IV. ENERGY CONSUMPTION

Finally, we discuss the energy consumption in the different steps described in Figure 4 for the application of one input current pulse. Input and learning pulses (steps 2 and 6 in Fig. 4) with an amplitude of tens of µA and a 300 ns width can be applied with CMOS spiking neurons with an energy consumption as low as 7 pJ, including the dissipation of the circuit [38]. In optimized small devices, switching can be achieved with shorter pulses (1-10 ns) of that amplitude, and therefore the energy consumption can be expected to be lower than 1 pJ per input pulse. The reconfigurable classifier based on N devices requires N learning pulses with an amplitude well below the amplitude of input pulses. We estimate that an amplitude below 20% of the amplitude of input pulses will be required to recognize patterns similar to those used to design the array before fabrication. In this scenario, the total energy can be estimated to be below ~ $(1+0.2·N)$ pJ. As will be seen in the following, this step dominates the energy consumption of the classifier.

The energy to read the devices configuration in STT-MRAM memories is lower than the energy of the writing process [42]. A CMOS circuit can easily detect a variation of voltage of 50 mV consuming 200 fJ per device (a value which decreases upon increasing the magnitude of the voltage variation), as we discussed in [6]. Other studies of STT MRAM memories have reported lower values of energy consumption, for instance 70 fJ per device in [43]. The binary vector describing the switching configuration can be transformed into a unique output (step 5 in Figure

4) through a set of AND/OR/etc gates or, for the system to be reconfigurable, with lookup tables (LUTs) like those used in field programmable gate arrays (FPGAs). An 8-inputs LUT based on STT-MRAM technology has an energy consumption of 10 fJ [41].

In summary, the energy consumption of the classifier is dominated by the energy required to apply input and learning pulses and, to a lesser extent, by the energy required to detect switching. The total energy consumption of the classifier can be estimated to be below ~(1+0.2·N) pJ per applied input, which is orders of magnitude lower than GPUs or the TrueNorth chip [2].

## V. CONCLUSION

Our experimental study shows that arrays of interconnected magnetic tunnel junctions as those used in STT-MRAM memories can learn to classify data encoded in the amplitude of input currents into categories through the spin torque driven magnetization switching output configuration. We demonstrate that a network based on three electrically connected junctions can learn to classify seven spoken vowels with a recognition rate of 96%, despite the intrinsic stochasticity of the switching process. This surpasses the performance of multilayer software neural networks with a similar number of trained parameters in the same task. These results open the path to the use of STT-MRAM memory arrays as neuromorphic reconfigurable classifiers.

**Acknowledgments**

We acknowledge Prof. J. Santamaría for fruitful discussions. This work was supported by Grant Nos. PID2020-116181RB-C33, PID2020-116181RB-C31, PGC2018-099422-A-I00 and CNS2022-136053 funded by MCIN/AEI/ 10.13039/501100011033, ERDF "A way of making Europe" and European Union "Next generation EU/PRTR"; and by Grant Nos. S2018/NMT-4321 (NanomagCOST-CM) and 2018-T1/IND-11935 (Atracción de Talento) funded by Comunidad de Madrid. A. Lopez was funded through the FPU Fellowship No. FPU20/02408. IMDEA-Nanociencia acknowledges support from the 'Severo Ochoa' Program for Centers of Excellence in R&D (CEX2020-001039-S).
**References**
1. G. Indiveri, S. C. Liu. Memory and Information Processing in Neuromorphic Systems. Proc. IEEE 103, 1379–1397 (2015).
2. P. A. Merolla, J. V. Arthur, R. Alvarez-icaza, A. S. Cassidy, J. Sawada, F. Akopyan, B. L. Jackson, N. Imam, C. Guo, Y. Nakamura et al. A million spiking-neuron integrated circuit with a scalable communication network and interface. Science 345, 668–673 (2014).
3. D. Querlioz, O. Bichler, A. F. Vincent, C. Gamrat. Bioinspired Programming of Memory Devices for Implementing an Inference Engine. Proc. IEEE 103, 1398–1416 (2015).
4. N. Locatelli, V. Cros, J. Grollier. Spin-torque building blocks. Nat. Mater. 13, 11–20 (2014).
5. J. Torrejon et al. Neuromorphic computing with nanoscale spintronic oscillators. Nature 547, 428–431 (2017).
6. M. Romera, P. Talatchian, S. Tsunegi, F. A. Araujo, V. Cros, P. Bortolotti, K. Yakushiji, A. Fukushima, H. Kubota, S. Yuasa et al. Vowel recognition with four coupled spin-torque nano-oscillators. Nature 563, 230–234 (2018).
7. A. Mizrahi, T. Hirtzlin, A. Fukushima, H. Kubota, S. Yuasa, J. Grollier, D. Querlioz. Neural-like computing with populations of superparamagnetic basis functions. Nat. Commun. 9, 1533 (2018).

8. G. Srinivasan, A. Sengupta, K. Roy. Magnetic Tunnel Junction Based Long-Term Short-Term Stochastic Synapse for a Spiking Neural Network with On-Chip STDP Learning. Sci. Rep. 6, 29545 (2016).

9. M. Romera, P. Talatchian, S. Tsunegi, K. Yakushiji, A. Fukushima, H. Kubota, S. Yuasa, V. Cros, P. Bortolotti, M. Ernoult et al. Binding events through the mutual synchronization of spintronic nano-neurons. Nat. Commun. 13, 883 (2022).

10. J. Grollier, D. Querlioz, K. Y. Camsari, S. Fukami, M. D. Stiles. Neuromorphic spintronics. Nat. Electron. 3, (2020).

11. P. Talatchian, M. Romera, F. Abreu Araujo, P. Bortolotti, V. Cros, D. Vodenicarevic, N. Locatelli, D. Querlioz, J. Grollier. Designing Large Arrays of Interacting Spin-Torque Nano-Oscillators for Microwave Information Processing. Phys. Rev. Appl. 13, 024073 (2020).

12. J. Zhou, J. Chen. Prospect of Spintronics in Neuromorphic Computing. Adv. Electron. Mater. 2100465 (2021).

13. C. M. Liyanagedera, A. Sengupta, A. Jaiswal, K. Roy. Stochastic Spiking Neural Networks Enabled by Magnetic Tunnel Junctions: From Nontelegraphic to Telegraphic Switching Regimes. Phys. Rev. Appl. 8, 064017 (2017).

14. T. Böhnert, Y. Rezaeiyan, M. S. Claro, L. Benetti, A. S. Jenkins, H. Farkhani, F. Moradi, R. Ferreira. Weighted spin torque nano-oscillator system for neuromorphic computing. Commun. Eng. 2, 65 (2023). https://doi.org/10.1038/s44172-023-00117-9

15. A. Ross, N. Leroux, A. De Riz, D. Marković, D. Sanz-Hernández, J. Trastoy, P. Bortolotti, D. Querlioz, L. Martins, L. Benetti, M. S. Claro, P. Anacleto, A. Schulman, T. Taris, J.-B. Begueret, S. Saïghi, A. S. Jenkins, R. Ferreira, A. F. Vincent, F. A. Mizrahi, J. Grollier. Nat. Nanotechnol. 18, 1273 (2023).

16. R. Khymyn, I. Lisenkov, J. Voorheis, O. Sulymenko, O. Prokopenko, V. Tiberkevich, J. Akerman, A. Slavin. Ultra-Fast Artificial Neuron: Generation of Picosecond-Duration Spikes in a Current-Driven Antiferromagnetic Auto-Oscillator. Sci. Rep. 8, 15727 (2018).

17. A. Kurenkov, S. DuttaGupta, C. Zhang, S. Fukami, Y. Horio, H. Ohno. Artificial Neuron and Synapse Realized in an Antiferromagnet/Ferromagnet Heterostructure Using Dynamics of Spin–Orbit Torque Switching. Adv. Mater. 31, 1900636 (2019).

18. M. Zahedinejad, A. A. Awad, S. Muralidhar, R. Khymyn, H. Fulara, H. Mazraati, M. Dvornik, J. Åkerman. Two-dimensional mutually synchronized spin Hall nano-oscillator arrays for neuromorphic computing. Nat. Nanotechnol. 15, 47–52 (2020).

19. Q. Yang, R. Mishra, Y. Cen, G. Shi, R. Sharma, X. Fong, H. Yang. Spintronic Integrate-Fire-Reset Neuron with Stochasticity for Neuromorphic Computing. Nano Lett. 22, 8437 (2022).

20. D. R. Rodrigues, R. Moukhader, Y. Luo, B. Fang, A. Pontlevy, A. Hamadeh, Z. Zeng, M. Carpentieri, G. Finocchio. Spintronic Hodgkin-Huxley-Analogue Neuron Implemented with a Single Magnetic Tunnel Junction. Phys. Rev. Applied 19, 064010 (2023).

21. S. Li, A. Du, Y. Wang, X. Wang, X. Zhang, H. Cheng, W. Cai, S. Lu, K. Cao, B. Pan, N. Lei, W. Kang, J. Liu, A. Fert, Z. Hou, W. Zhao. Science Bulletin 67, 691–699 (2022).

22. J. Von Kim. Spin-Torque Oscillators. Solid State Physics - Advances in Research and Applications vol. 63 (2012).

23. D. Vodenicarevic, N. Locatelli, A. Mizrahi, J. S. Friedman, A. F. Vincent, M. Romera, A. Fukushima, K. Yakushiji, H. Kubota, S. Yuasa, S. Tiwari, J. Grollier, D. Querlioz. Low-Energy Truly Random Number Generation with Superparamagnetic Tunnel Junctions for Unconventional Computing. Phys. Rev. Applied 8, 054045 (2017).

24. J. Cai, B. Fang, L. Zhang, W. Lv, B. Zhang, T. Zhou, G. Finocchio, Z. Zeng. Voltage-Controlled Spintronic Stochastic Neuron Based on a Magnetic Tunnel Junction. Phys. Rev. Applied 11, 034015 (2019).

25. K. Y. Camsari, B. M. Sutton, S. Datta. Appl. Phys. Rev. 2019, 6, 011305.

26. A. Sengupta, P. Panda, P. Wijesinghe, Y. Kim, K. Roy. Magnetic tunnel junction mimics stochastic cortical spiking neurons. Sci. Rep. 6, 1–9 (2016).

27. A. Mizrahi, N. Locatelli, J. Grollier, D. Querlioz. Phys. Rev. B 94, 054419 (2016).

28. J. Grollier, V. Cros, A. Hamzic, J. M. George, H. Jaffrès, A. Fert, G. Faini, J. Ben Youssef, H. Legall. Spin-polarized current induced switching in Co/Cu/Co pillars. Appl. Phys. Lett. 78, 3663–3665 (2001).

29. A. F. Vincent, J. Larroque, N. Locatelli, N. Ben Romdhane, O. Bichler, C. Gamrat, W. Zhao, J.-O. Klein, S. Galdin-Retailleau, D. Querlioz. Spin-transfer torque magnetic memory as a stochastic memristive synapse for neuromorphic systems. IEEE Trans. Biomed. Circuits Syst. 9, 166–174 (2015).

30. E. Raymenants, A. Vaysset, D. Wan, M. Manfrini, O. Zografos, O. Bultynck, J. Doevenspeck, M. Heyns, I. P. Radu, T. Devolder. Chain of magnetic tunnel junctions as a spintronic memristor. J. Appl. Phys. 124 (2018).

31. P. Rzeszut, J. Chęciński, I. Brzozowski, S. Ziętek, W. Skowroński, T. Stobiecki. Multi-state MRAM cells for hardware neuromorphic computing. Sci Rep 12, 7178 (2022).

32. S. Jung, H. Lee, S. Myung, H. Kim, S. K. Yoon, S. W. Kwon, Y. Ju, M. Kim, W. Yi, S. Han, B. Kwon, B. Seo, K. Lee, G. H. Koh, K. Lee, Y. Song, C. Choi, D. Ham, S. J. Kim. A crossbar array of magnetoresistive memory devices for in-memory computing. Nature 601, 211–216 (2022).

33. J. D. Costa, S. Serrano-Guisan, B. Lacoste, A. S. Jenkins, T. Böhnert, M. Tarequzzaman, J. Borme, F. L. Deepak, E. Paz, J. Ventura, R. Ferreira, P. P. Freitas. High power and low critical current density spin transfer torque nano-oscillators using MgO barriers with intermediate thickness. Sci. Rep. 7, 7237 (2017).

34. See Supplemental Material at [URL will be inserted by publisher], which includes Ref. [44], for details on the magneto-transport properties of the magnetic tunnel junctions, the database and the inputs applied to the experimental network, the real-time learning procedure, the automatic detection of magnetization switching and the multilayer perceptron simulations.

35. M. Gajek, J. J. Nowak, J. Z. Sun, P. L. Trouilloud, E. J. O'Sullivan, D. W. Abraham, M. C. Gaidis, G. Hu, S. Brown, Y. Zhu, R. P. Robertazzi, W. J. Gallagher, D. C. Worledge. Spin torque switching of 20 nm magnetic tunnel junctions with perpendicular anisotropy. Appl. Phys. Lett. 100, 132408 (2012).

36. S. B. Furber, D. R. Lester, L. A. Plana, J. D. Garside, E. Painkras, S. Temple, A. D. Brown. Overview of the SpiNNaker System Architecture. IEEE Trans. Comput. 62, 2454–2467 (2013).

37. Y. P. Lin, C. H. Bennett, T. Cabaret, D. Vodenicarevic, D. Chabi, D. Querlioz, B. Jousselme, V. Derycke, J. O. Klein. Physical Realization of a Supervised Learning System Built with Organic Memristive Synapses. Sci. Rep. 6, 31932 (2016).

38. P. Livi, G. Indiveri. A current-mode conductance-based silicon neuron for Address-Event neuromorphic systems. 2898–2901 (2009).

39. K. Watanabe, B. Jinnai, S. Fukami, H. Sato, H. Ohno. Shape anisotropy revisited in single-digit nanometer magnetic tunnel junctions. Nat. Commun. 5–10 (2018). doi:10.1038/s41467-018-03003-7.

40. H. Park, R. Dorrance, A. Amin, F. Ren, D. Marković, C. K. Ken Yang. Analysis of STT-RAM Cell Design with Multiple MTJs Per Access. 2011 IEEE/ACM International Symposium on Nanoscale Architectures, San Diego, CA, USA, 2011, pp. 53-58.

41. K. Jo, K. Cho, H. Yoon. Variation-Tolerant and Low Power Look-Up Table (LUT) Using Spin-Torque Transfer Magnetic RAM for Non-volatile Field Programmable Gate Array (FPGA). 2016 International SoC Design Conference (ISOCC), Jeju, Korea (South), 2016, pp. 101-102. doi:10.1109/ISOCC.2016.7799753.

42. K. Emre. Evaluating STT-RAM as an Energy-Efficient Main Memory Alternative. 2013 IEEE International Symposium on Performance Analysis of Systems and Software (ISPASS), 256–267 (2013).

43. B. Song, T. Na, J. Kim, J. P. Kim, S. H. Kang, S. O. Jung. Latch Offset Cancellation Sense Amplifier for Deep Submicrometer STT-RAM. IEEE Transactions on Circuits and Systems I: Regular Papers, vol. 62, no. 7, pp. 1776-1784, July 2015. doi:10.1109/TCSI.2015.2427931.
44. J. Hillenbrand, L. A. Getty, K. Wheeler, M. J. Clark. Acoustic characteristics of American English vowels. J. Acoust. Soc. Am. 97, 3099–3111 (1994).